\documentclass[12pt,dvips]{article}
\usepackage{graphicx}
\usepackage{amsmath}
\usepackage[psamsfonts]{amssymb}
\usepackage{amsthm}
\usepackage{euscript}

\usepackage{latexsym}

\renewcommand{\theequation}{\arabic{section}.\arabic{equation}}
\renewcommand{\(}{\begin{equation}}
\renewcommand{\)}{end{equation} \vspace{-.05in}\linebreak}

\newcounter{saveeqn}
\newcounter{savealpheqn}

\newcommand{\alpheqn}{\setcounter{saveeqn}{\value{equation}}%
 \stepcounter{saveeqn}\setcounter{equation}{0}%
 \renewcommand{\theequation}{\mbox{\arabic{section}.\arabic{saveeqn}\alph{equation}}}
 \renewcommand{\)}{\end{equation}}}
\def\part#1{\frac{\partial}{\partial{#1}}}%

\def\group#1{\refstepcounter{equation}\setcounter{saveeqn}{\value{equation}}%
\label{#1}\setcounter{equation}{0}%
\renewcommand{\theequation}{\mbox{\arabic{section}.\arabic{saveeqn}\alph{equation}}}%
\renewcommand{\)}{\end{equation}}}

\newcommand{\reseteqn}{\setcounter{equation}{\value{saveeqn}}%
 \renewcommand{\theequation}{\arabic{section}.\arabic{equation}}%
 \renewcommand{\)}{\end{equation}}}

\newcommand{\aalpheqn}{\setcounter{saveeqn}{\value{equation}}%
 \stepcounter{saveeqn}\setcounter{equation}{0}%
 \renewcommand{\theequation}{\mbox{\Alph{subsection}.\arabic{saveeqn}\alph{equation}}}
  \renewcommand{\)}{\end{equation}}}

\newcommand{\areseteqn}{\setcounter{equation}{\value{saveeqn}}%
 \renewcommand{\theequation}{\Alph{subsection}.\arabic{equation}}%
 \renewcommand{\)}{\end{equation}}}

\renewcommand{\thefootnote}{\alph{footnote}}
\renewcommand{\(}{\begin{equation}}
\renewcommand{\)}{\end{equation}}
\newcommand{\ba}{\begin{eqnarray}}
\newcommand{\ea}{\end{eqnarray}}

\newcommand{\bp}{\mathop{\vtop{\ialign{##\crcr
  $\hfil\displaystyle{}\hfil$\crcr\noalign{\kern-13pt\nointerlineskip}
  \BIG{(}\hskip0pt\crcr\noalign{\kern3pt}}}}}
\newcommand{\cbp}{\mathop{\vtop{\ialign{##\crcr
  $\hfil\displaystyle{}\hfil$\crcr\noalign{\kern-13pt\nointerlineskip}
  \BIG{)}\hskip0pt\crcr\noalign{\kern3pt}}}}}
\newcommand{\pa}{\mathop{\vtop{\ialign{##\crcr
  $\hfil\displaystyle{\oplus}\hfil$\crcr\noalign{\kern+1pt\nointerlineskip}
  \hspace{.08in}$^{\alpha=0}$\hskip6pt\crcr\noalign{\kern3pt}}}}}
\renewcommand{\sp}{,\hspace{.3in}}

\newcommand{\R}{\ensuremath{\mathbb R}}

\newcommand{\Z}{\ensuremath{\mathbb Z}}

\def\rpf{\ensuremath{\mathbf{RP}^4}}
\def\rpt{\ensuremath{\mathbf{RP}^2}}
\def\rpn{\ensuremath{\mathbf{RP}^n}}
\def\of{O4${}^-$}
\def\ofh{O4${}^+$}
\def\ofpm{O4${}^\pm$}
\def\ofg{$\widetilde{\textup{O4}}^-$}
\def\ofgh{$\widetilde{\textup{O4}}^+$}
\def\ofgpm{$\widetilde{\textup{O4}}^\pm$}

\newcommand{\beq}{\begin{equation}}
\newcommand{\eeq}{\end{equation}}

\numberwithin{equation}{section}

\catcode`\@=11
\def\vereq#1#2{\lower3pt\vbox{\baselineskip1.5pt \lineskip1.5pt
\ialign{$\m@th#1\hfill##\hfil$\crcr#2\crcr\sim\crcr}}}
\catcode`\@=12

\makeatletter
\newcommand\tabcaption{\def\@captype{table}\caption}
\makeatother
\renewcommand{\(}{\begin{equation}}
\renewcommand{\)}{\end{equation}}

\begin{document}

\begin{titlepage}
\begin{flushright}
IFUP-TH/2003/16

hep-th/0304098
\end{flushright}

\vspace{2em}
\def\thefootnote{\fnsymbol{footnote}}

\begin{center}
{\Large  A Torsion Correction to the RR 4-Form Fieldstrength}
\end{center}
\vspace{1em}

\begin{center}
Jarah Evslin\footnote{E-Mail: jarah@df.unipi.it} 
\end{center}

\begin{center}
\vspace{1em}
{\em INFN Sezione di Pisa\\
     Universita di Pisa\\
     Via Buonarroti, 2, Ed. C,\\
     56127 Pisa, Italy}\\
\end{center}

\vspace{3em}
\begin{abstract}
\noindent

\end{abstract}
The shifted quantization condition of the M-theory 4-form $G_4$ is well-known.  The most naive generalization to type IIA string theory fails, an orientifold fourplane counterexample was found by Hori in hep-th/9805141.  In this note we use D2-brane anomaly cancellation to find the corresponding shifted quantitization condition in IIA.  Our analysis is consistent with the known O4-plane tensions if we include a torsion correction to the usual construction of $G_4$ from $C_3$, $B$ and $G_2$.  The resulting Bianchi identities enforce that RR fluxes lift to K-theory classes.

\vfill
10 April, 2003

\end{titlepage}
\setcounter{footnote}{0} 
\renewcommand{\thefootnote}{\arabic{footnote}}

\pagebreak
\renewcommand{\thepage}{\arabic{page}}
\pagebreak 

\section{Introduction}

A quantum theory is consistent if all of the quantities that it computes are well-defined.  In particular, the partition function on the worldvolume of each kind of object, including all corrections from inflows, must admit some consistent definition.  A consistent definition is one which depends only on the position of the object in parameter space, and so is required to be invariant under the translation about any loop in parameter space.  The partition function is the exponential of the effective action, and so we require that the effective action shift only by an integer as one journeys around such a loop.  Often a version of Stoke's theorem may be applied to reexpress this shift as an integral of a derivative of the effective action over the loop, and so consistency demands that this quantity be integral.  Such a condition is called a Dirac quantization condition.

In Ref.~\cite{FluxQuant} Witten applied this argument to a closed M2-brane in M-theory.  Two terms in the M2-brane's worldvolume action were considered, the integral of the M-theory 3-form $C_3$ and also the worldvolume fermion term.  Both of these terms could potentially be ill-defined, the first because $C_3$ is not gauge-invariant, and the second because while the path integral of the fermions is real, its sign is not canonically defined and so potentially may change as one encircles such a loop.  The total shift of the effective action was calculated to be
\begin{equation}
S\longrightarrow S + \int_M (G_4 - \pi\lambda)
\end{equation}
where $\lambda$ is half of the first Pontrjagin class of the tangent bundle of spacetime and $M$ is the four-cycle swept out by the membrane's three-volume during its loop through parameter space. 

This leads to the Dirac quantization condition
\begin{equation}
\int_M \frac{G_4}{2\pi} - \frac{\lambda}{2}\in \Z.
\end{equation}
Here $M$ is any 4-cycle in the spacetime $Y^{11}$, as all 4-cycles are loops of embeddings of membranes.  As $M$ may be any class in $H_4(Y^{11})$, it was hinted in \cite{FluxQuant} that
\begin{equation}
\frac{G_4}{2\pi} + w_4 \in H^4(Y^{11};\Z) \label{mquant}
\end{equation}
where $w_4$ is the 4th Stiefel Whitney class of the tangent bundle of spacetime.  In the orientable case $w_4$ (normalized so that each component is zero or one half) agrees with $\lambda/2$ modulo one, while in the nonorientable case only $w_4$ continues to be well-defined.  $w_4$ has no natural lift to integral cohomology, however the choice of lift does not alter the quantization condition in (\ref{mquant}).  Thus the M-theory 4-form $G_4$ is classified by integral cohomology, but with a possible half-integer shift corresponding to a $\Z_2$ extension \cite{Sugimoto}.

It has been widely speculated that this shifted quantization condition for $G_4$ extends to type IIA string theory.  However an apparent counterexample has been known for several years \cite{Kentaro}.  As reviewed in Section~\ref{o4sec}, there are four types of orientifold 4-plane, distinguished by discrete torsion $H$ and $G_2$ fluxes on the \rpf's\ that link them.  Due to the action of the orientifold on the field $G_4$, instead of inhabiting $\Z$-cohomology, 4-forms are classified by $\tilde{H}^4$, the cohomology group with coefficients in the $\Z_2$-twisted sheaf.\footnote{One may choose to dismiss the counterexample because the twisting may affect the shifted quantization condition.  However this appears to be the canonical example of the shift in the literature, and so we feel that it merits a closer examination.}  This group is $\Z$, and so a naive application of the twisted quantization, using the fact that $w_4=1/2$ for \rpf, would lead to the conclusion\footnote{From this equation on we will absorb a factor of $1/2\pi$ into our fluxes.} that $\int G_4\in\Z+\frac{1}{2}$.  However the $G_4$ fluxes on the \rpf\ linking the O4-plane are known and for one type of discrete torsion the flux vanishes, violating the shifted quantization condition.

Such a failure might have been expected, after all, in type IIA there are several notions of four-form field strength, any of which might be quantized.  To learn which is quantized, in Sec.~\ref{d2sec}, we will repeat the M2-brane quantization argument with D2-branes, whose worldvolume action has an extra $B\wedge C_1$ with respect to that of the M2.  Using this new quantity, we find that instead of one choice of discrete torsions violating the quantization condition, two choices now appear to be anomalous.

The non-anomolous discrete torsion orientifolds may be changed into the troublesome orientifolds by aborbing a one-half D6-brane, and so in Sec.~\ref{d6sec} we turn our attention to the worldvolume gauge theory of this D6-brane.  We see that the quantization of the gauge field on this D6 is shifted by a half-integer as a result of the nonvanishing $w_2$ of its normal bundle.  This in turn induces a half-integral D4-charge, which leads to a torsion correction to the known construction of $G_4$.  The resulting construction
\begin{equation}
G_4=dC_4+(sq^2+B\cup)G_2
\end{equation}
yields a quantization condition that is satisfied by all four varieties of O4-plane.  Of course, like $B$, the $sq^2$ term defies a natural lift to integral cohomology.  In the spirit of \cite{MMS}, we interpret this as an ambiguity of the lift of $dC_4$ (whose lift to cohomology was already not naturally-defined, it being better described as a K-theory class \cite{MM,Sugimoto}) related to the decay of topologically nontrivial D4-branes via a mortal NS5 or D6-brane. 

Such a correction leads to a modification of the Bianchi identity
\begin{equation}
dG_4=(Sq^3+H\cup)G_2=d_3 G_2 \label{mod}
\end{equation}
where $d_3$ is a differential in the Atiyah-Hirzebruch spectral sequence (AHSS).  This modified Bianchi identity enforces that, in the absence of D4-branes, $G_2$ lifts to a K-theory class.  This is the converse of the claim of Freed and Hopkins in Ref.~\cite{FH}.  Of course $Sq^3$ annihilates any two-form, but later we will argue that in T-dual cases the spectra of orientifold planes are consistent with the extension of Eq.~(\ref{mod}) to other RR fluxes.

\section{Orientifold 4-Plane Review} \label{o4sec}
This note is an investigation of the consistency of D2-branes in the presence of the known variants of O4-planes.  And so we will begin by reviewing a few facts about these planes which were largely derived in Ref.~\cite{Kentaro} and have been reviewed in Ref.~\cite{HK}.  Consider IIA string theory on $\R^{1,9}$ with basis $x^i$, $0\leq i\leq 9$.  Now gauge this theory by the $\Z_2$ symmetry
\begin{equation}
x^{i\leq 4}\leftrightarrow x^{i}\sp
x^{i\geq 5}\leftrightarrow -x^{i}\sp
B\leftrightarrow -B\sp
C_1\leftrightarrow C_1\sp
C_3\leftrightarrow -C_3
\end{equation}
along with $(-1)^{F_L}$ on the worldsheets of all F-strings.  The O4-plane is the fixed plane $x^{i\geq 5}=0$.  In the covering space, $\R^{1,9}$, it is linked by an 4-sphere, which in the quotient space descends to \rpf.  

Fluxes which are fixed by the projection are then classified by the cohomology of \rpf\ with $\Z$ coefficients $H^*(\rpf)$, while those which are negated by the projection are classified by cohomology with coefficients in $\Z$ twisted by the $\Z_2$ orientation bundle, $\tilde{H}^*(\rpf)$.  In particular
\begin{equation}
H\in \tilde{H}^3(\rpf)=\Z_2\sp
dC_1\in H^2(\rpf)=\Z_2\sp
dC_3\in \tilde{H}^4(\rpf)=\Z.
\end{equation}
 We will consider massless IIA, and so $G_2=dC_1$.  The integral of $dC_3$ measures the D4-brane charge of the plane, and so we see that any number of D4-branes may be linked by a given $\rpf$.  The two $\Z_2$ classes yield four variations of O4-plane, which we now list along with their tensions $G_4$: 

\vspace{.3in}
\hspace{2in}
\begin{tabular}{c|c|c|c}
Type&$H$&$G_2$&$G_4$\\ \hline
\of&0&0&-1/2\\
\ofh&1/2&0&1/2\\
\ofg&0&1/2&0\\
\ofgh&1/2&1/2&1/2\\
\multicolumn{4}{c}{}\\
\multicolumn{4}{c}{Table 1: O4-Planes}
\end{tabular}

\vspace{.3in}

\noindent
$w_4(\rpf)=1/2$ and so, as was first observed in Ref.~\cite{Kentaro}, the \ofg-plane violates the shifted quantization condition $G_4\in\Z+1/2$.

\section{The D2-Brane Anomaly} \label{d2sec}
However the D2-brane partition function is not the same as the M2-brane partition function\footnote{More precisely, the partition functions are equal, but the definitions of the fluxes are different in IIA and M-theory.  The M-theory flux $C_3$ decomposes into both $C_3$ and $B$ in IIA.}, and so it is far from obvious that same quantization condition applies to the D2 as the M2.  For example the D2-brane partition function contains an extra term $B\wedge C_1$, whose variation is $d(B\wedge C_1)$.  The D2, like the M2, couples electrically to $C_3$ and so again there is a variation of $dC_3$, however in IIA $dC_3\neq G_4$.  Therefore, as the D2-brane's path sweeps out a loop, the partition function changes by
\begin{equation}
e^{iS}\longrightarrow e^{iS + 2\pi i\int_M (w_4 + dC_3 + d(B\wedge C_1))}.
\end{equation}
The shifted quantization condition for type IIA string theory on $X^{10}$ is then
\begin{equation}
dC_3+d(B\wedge C_1) + w_4 \in \tilde{H}^4(X^{10}). 
\end{equation}
The 4th Stiefel-Whitney class $w_4$ always contributes $1/2$ for any type of O4-plane.  

$B$ and $C_1$ are valued in cohomology with $U(1)$-valued coefficients, and so we will need to choose a lift to $\Z$-valued cohomology.  There is no canonical choice for such a lift, however fortunately in the present case the choice of lift is inconsequential.  The reason for this is that on \rpf\ the only allowed values of $B$ and $C_1$ near the O4 are $0$ and $1/2$, and so a different integral lift corresponds to a shift by an even multiple of the fundamental unit.  However this leads to a shift in the 4-cohomology by an even multiple of the fundamental unit, $1/2$, which is integral and so does not affect whether a configuration satisfies the quantization condition.  

In fact, the exterior derivative here is the Bockstein homomorphism $\beta$ from $U(1)$ to $\Z$ cohomology that appears in the exact sequences
\begin{equation}
0=\tilde{H}^2(\rpf,\R)\longrightarrow\tilde{H}^2(\rpf,U(1))=\Z_2\stackrel{\beta}{\longrightarrow}\tilde{H}^3(\rpf,\Z)=\Z_2
\end{equation}
and
\begin{equation}
0=H^1(\rpf,\R)\longrightarrow H^1(\rpf,U(1))=\Z_2\stackrel{\beta}{\longrightarrow} H^2(\rpf,\Z)=\Z_2.
\end{equation}
This means that $d$ takes $0$ to $0$ and $1/2$ to $1/2$, which implies the last equality in 
\begin{equation}
d(B\wedge C_1)=dB\wedge C_1+B\wedge dC_1=2B\wedge C_1.
\end{equation}
Again we see that this term contributes an even muliple of $1/2$, which is an integer and so does not affect whether the quantization condition is violated.  It is quite fortunate that this term vanishes, as we do not know whether in general it is cancelled by a bulk contribution as in Ref.~\cite{Taylor}.

Combining these facts, the O4-plane is consistent with D2-brane anomaly cancellation if
\begin{equation}
dC_3\in \Z+\frac{1}{2}.
\end{equation}
Now we can compare this quantization condition with the known spectrum of O4-planes, calculating $dC_3$ via the supergravity relation
\begin{equation}
G_4=dC_3+B\wedge G_2. \label{old}
\end{equation}
Notice that this quantization condition only differs from our old quantization condition when $B\wedge G_2\neq 0$, that is, for the \ofgh-plane.  Thus instead of curing the failure of the quantization condition $G_4\in\Z+1/2$ for the \ofg, our correction has led to the failure of quantization for a second type of O4!

\section{A Correction to $G_4$} \label{d6sec}

\begin{figure}[ht]
  \centering \includegraphics[width=3in]{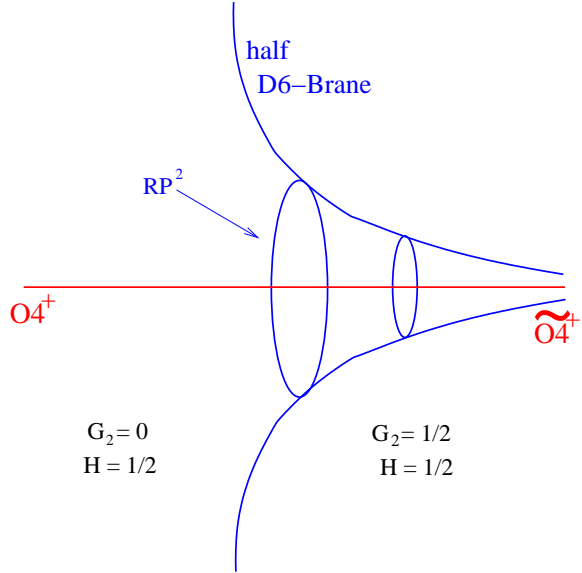}
\caption{On the left is an \ofh-plane, which is linked by an \rpf\ supporting half a unit of $H$ flux and no $G_2$ flux.  The intersection of the plane with a half D6-brane is a domain wall in the worldvolume of the plane.   The half D6-brane separates regions of $G_2=0$ and $G_2=1/2$, and so on the right side of the half D6 the orientifold plane is an \ofgh.  At vanishing string coupling, the D6 is flat, but for any finite string coupling it is funnel-shaped and the intersection with the orientifold plane is at infinity.} \label{d6fig}
\end{figure}

Two O4-planes (\ofpm) satisfy the quantization condition, while the other two (\ofgpm) appear not to.  The difference between these two pairs of planes is that the $G_2$ flux is turned off in the well-behaved \ofpm, but is turned on in the problematic \ofgpm.  One implication is that an \ofpm-plane turns into an \ofgpm when it absorbs a D6-brane \cite{Kentaro}, as drawn in Figure \ref{d6fig}.  Thus we are motivated to more carefully examine the D6-absorption process to look for corrections to (\ref{old}).

This process is often described in the strict limit $g_s\rightarrow 0$ because in the covering space the D6-brane becomes a flat $\R^7$\ extending along directions $x^{i}$ with $1\leq i \leq 7$ and existing only in the instant $x^0=0$.   Taking the $\Z_2$ quotient of this configuration, where the D6 is its own mirror image, leads to a $1/2$ D6-brane which intersects the O4-plane at a fixed timeslice, at which point the O4 changes from a \ofpm\ to a \ofgpm.  This intersection is singular, but the singularity and in fact the intersection itself are removed at finite $g_s$.  In this case the D6 can again be taken to appear at time $x^0=0$.  However instead of completing its life in this same instant, at each moment $x^0=t>0$ the D6 is geometrically $\R^4\times\rpt$ with the $\R^4$ extending along directions $x^i$, $1\leq i\leq 4$ and the $\rpt$ the $\Z_2$ quotient of the sphere of radius $1/t$ which is extended in directions $x^5$, $x^6$\ and $x^7$\ and is centered about the O4-plane.  This system is a quotient of a T-dual of the D3-D5 system considered in \cite{HanWit}.  Our choice of time direction makes the D6-brane into an S-brane \cite{S-branes} far away from the O4-plane.  If one wishes to avoid S-branes, one may change the geometry far away, as the argument below concerns only the region near the O4-plane.  Alternately, one may choose a different time direction.

At every moment the O4-plane is of type \ofpm.  However if one probes the system at any fixed lengthscale $d$ one concludes that the \ofpm\ absorbs the D6-brane and changes to a \ofgpm-plane at time $x^0=1/d$.  The \ofg\ plane has a higher tension than its \of\ counterpart, corresponding to a higher 4-brane charge.  Here we are using a convention typical in the orientifold literature, and refering to the 4-brane charge as the integral of $G_4$ over a link, rather than the integral of $dC_3$.  The second quantity is not invariant under supergravity gauge transformations and so cannot compute a classical observable like a tension.  In fact, the second quantity may not change upon intersection with a D6-brane as that would lead to $dddC_3\neq 0$ at the point of intersection.  Therefore at the point of intersection of the O4-plane with the D6, the cohomology class of $G_4$ over a linking \rpf\ jumps by the D4 charge of the D6-brane, while the class of $dC_3$ is fixed.  This means that $G_4-dC_3$ jumps by the D4 charge of the D6, and so computing this charge will allow us to test (\ref{old}).

The D6-brane hosts a $U(1)$ gauge fieldstrength $F$, and its D4-brane charge on any two-cycle is the integral of $B+F$ on that two-cycle.  The integral of $B$, as described above, depends on the type of orientifold plane we consider.  The \of\ yields $B=0$, while the \ofh\ carries $B=1/2$.  The integral of $F$ is determined by the requirement that U(1)-charged fermions may exist on the D6-brane worldvolume.  The fermions inhabit a section of the bundle $N^{1/2}\otimes G$ where $N$ is the normal bundle to $\rpt$ and $G$ is the $U(1)$ gauge bundle.  If $N^{1/2}$ is itself a bundle, that is, if it satisfies the 3-way transition function identity $f_{ij}f_{jk}f_{ki}=1$, then $G$ must also be a bundle and so the integral of $F$ is an integer.  If on the other hand $N^{1/2}$ is not a bundle, then because $N$ is a bundle some set of transition functions must satisfy $f_{ij}f_{jk}f_{ki}=-1$.  To ensure that the tensor with $G$ is a bundle, $G$ must also not be a bundle but also have such transition functions.  In this case, the integral of $F$, the first Chern class of $G$, must be a half-integer.  $N^{1/2}$ is a bundle precisely if $w_2(N)=0$, and so
\begin{equation}
\int_{\rpt} F \in \Z+w_2(N) \label{fshift}
\end{equation}
where the second Stiefel-Whitney class $w_2\in H^2(\rpt)=\Z_2$ is normalized to be $0$ or $1/2$.  Such a shift in the presence of a nonvanishing $w_2(N)$ was previously seen in Ref.~\cite{TopEff}.

To calculate $w_2(N)$, we use the fact that the total Stiefel-Whitney class of \rpn\ is $(1+w_1)^{n+1}$ and the behavior of Stiefel-Whitney classes under direct sum to compute
\begin{equation}
w_1(N)=w_1(\rpf)-w_1(\rpt)=1/2-1/2=0
\end{equation}
from which it follows that
\begin{equation}
0=w_2(T\rpf)=w_2(T\rpt\oplus N)=w_2(T\rpt)+w_1(T\rpt)w_1(N)+w_2(N)=1/2+0+w_2(N).
\end{equation}
Thus we conclude that 
\begin{equation}
w_2(N)=1/2
\end{equation}
which yields the shifted quantization condition for the fieldstrength $F$.

We have learned that the worldvolume fieldstrength obeys the shifted quantization condition, while $B$ obeys the shifted quantization for the \ofh-plane but not the \of.  The D4-charge of the D6 is the sum of the contributions from $B$ and $F$.  This sum is not determined from the above data, as we are free to shift $F$ by any integer by adding a tube of magnetic flux to our configuration, and the lift of $B$ to $\Z$-cohomology is ill-defined, again allowing a shift by an arbitrary integer.  In fact, only the sum $B+F$ is gauge-invariant.  However possible integral shifts are irrelevant to the question at hand, the quantization condition of $B+F$.  Adding the two pieces together we find that the D4-charge $G_4$ is shifted for the \of-plane but not for the \ofh.  This is consistent with Table 1, where we saw that inclusion of the $G_2$ flux only changes the tension of the \of-plane.  

In fact, we could have simply taken the tension difference from the table and avoided the short computation above, however the way the shift came about in the above computation yields a formula for the shift, via (\ref{fshift}), that appears to contribute to the general case.  The difference between $G_4$ and $dC_3$, the D4-charge of the D6, was given up to an integer shift by 
\begin{equation}
G_4=dC_3+(F+B)\cup G_2=dC_3+(w_2(N(PD(dG_2))+B)\cup G_2. \label{exact}
\end{equation}
The $w_2$ correction may be approximated by a Steenrod square, yielding our main result
\begin{equation}
G_4=dC_3+(sq^2+B\cup)G_2.
\end{equation}
Of course, T-dualities may well lead to the natural generalization
\begin{equation}
G_p=dC_{p-1}+(sq^2+B\cup)G_{p-2}. \label{main}
\end{equation}
In fact, for every dimension of O$p$-plane the $G_{p-2}$ discrete torsion changes the tension by $1/2$ when $B=0$ and does not change the tension when $B=1/2$, in accordance with the fact that $w_2+B$ is equal to $1/2$ and $0$ in the two cases\footnote{To see this, follow the same calculation as above.  Notice that while $w_1$ of the tangent bundles may or may not vanish, they always agree and so $w_1(N)=0$ in every case.  Similarly $w_2$ of the tangent bundles always disagree and so $w_2(N)=1/2$ as in the O4 case.}.

In the classical limit torsion terms vanish and Eq.~(\ref{main}) reduces to the usual supergravity construction of the gauge invariant fieldstrength (\ref{old}).

\subsection{The Corrected Bianchi Identity}
The Bianchi identity is derived by requiring that $d^2$ annihilate the $(p-1)$-form connection $C_{p-1}$.  The torsion correction to $G_p$ then yields a torsion correction to the Bianchi identity
\begin{equation}
0=ddC_{p-1}=d(G_p-(sq^2+B\cup)G_{p-2})=dG_p-(Sq^3+H\cup) G_{p-2}.
\end{equation}
We may rewrite this in terms of the differentials of the AHSS
\begin{equation}
d_1G_p=d_3G_{p-2}.
\end{equation}
In the absence of D$(8-p)$-brane charge the left hand side vanishes and so $G_{p-2}$ is annihilated by $d_3$.  In type IIA this enforces that $G_{p-2}$ lifts to a [twisted] K-theory class.  

In type IIB, if there is $H$-flux, one must also enforce that $d_5$ annihilates $G_{p-2}$, requiring a higher order correction.  However it has been conjectured that the vanishing of the Freed-Witten anomaly is a sufficient condition for RR fields to lift to K-theory.  In this case, the Bianchi identity resulting from Eq.~(\ref{exact}) forces RR fields to lift to K-theory classes also in IIB.  However when we approximated $w_2$ by a Steenrod square we dropped a higher order correction which would have contributed to $d_5$ \cite{MMS}, and so we ruined our chances of enforcing the lift to K-theory in the modified Bianchi identity.  However, the fact that the correction in Eq.~(\ref{main}) enforces the K-theory classification in IIA and also in IIB in the absence of $H$-flux lends credibility to the conjecture that it applies to cases other than the O4-plane example considered here.  

\section{Conclusion}
One variety of orientifold O4-plane is inconsistent with the $G_4$ flux quantization condition that naively generalizes the known condition from M-theory.  The M-theory quantization condition is derived from an M2-brane worldvolume global anomaly, and we argue that in IIA this calculation differs in two ways.  First, in IIA the exterior derivative of $C_3$ is not equal to $G_4$.  Second, the corresponding D2-brane anomaly also receives a correction from the worldvolume action term $B\wedge C_1$, which has no analog in the M-theory case.  Combining these corrections we find that the quantization condition now fails for two varieties of O4-plane.

We note that these two varieties are the ones with nontrivial $G_2$ flux, and so study the process of D6-brane absorption, which toggles the offending $G_2$-flux.  We find that the D6-brane worldvolume $U(1)$ gauge fieldstrength obeys a shifted quantization condition due to the nontrivial topology of the orientifold.  This leads to a shift in the contribution of the D6-brane to the $G_4$ flux and in fact is necessary to obtain agreement with the known orientifold tensions.  We conjecture that this shift yields a torsion correction $sq^2G_2$ to the known supergravity construction of the gauge invariant fieldstrength $G_4=dC_3+B\wedge G_2$.  The shifted quantization condition (\ref{main}) is satisfied by all varieties of O4-plane.  As an additional check on the shift, we observe that it induces precisely the required shift on the Bianchi identities to enforce that RR-fluxes lift to twisted K-theory classes in the absence of D-branes.

\noindent 
{\bf Acknowledgements}

\noindent
I'd like to thank Eric Sharpe for his comments on an earlier draft and the INFN Sezione di Pisa for hiring me.
\noindent

\bibliographystyle{ieeetr} 
\bibliography{o4}
\end{document}